%% file: wadden.tex
\def\year{2021}\relax
\documentclass[letterpaper]{article} 
\usepackage{aaai21}  
\usepackage{times}  
\usepackage{helvet} 
\usepackage{courier}  
\usepackage[hyphens]{url}  
\usepackage{graphicx} 
\usepackage{natbib}  
\usepackage{caption} 
\usepackage{hyperref}

\usepackage{csvsimple}
\usepackage{subcaption}
\usepackage{listings}
\usepackage{lipsum,booktabs}
\usepackage{tabularx}
\usepackage{enumitem}
\usepackage{dialogue}
\usepackage{xcolor}
\usepackage{booktabs}

\newcommand{\xhdr}[1]{\vspace{1.7mm}\noindent{{\bf #1.}}}
\newcommand{\xhdrq}[1]{\vspace{1.7mm}\noindent{{\bf #1?}}}

\newcommand{\nmod}{\textsc{NoMod}}
\newcommand{\ymod}{\textsc{Mod}}

\newcommand*{\typewriter}{\fontfamily{lmtt}\selectfont}
\DeclareTextFontCommand{\texttypewriter}{\typewriter}

\newcommand{\user}[1]{\texttypewriter{#1}}
\newcommand{\moderator}[1]{\textit{#1}}

\newcommand{\userq}[1]{\speak{U} \texttypewriter{#1}}
\newcommand{\moderatorq}[1]{\speak{M} \textit{#1}}
\newcommand{\uone}[1]{\speak{U1} \texttypewriter{#1}}
\newcommand{\utwo}[1]{\speak{U2} \textit{#1}}

\input{gww-chars.tex}

\title{The Effect of Moderation on Online Mental Health Conversations}

\author {
  David Wadden, Tal August, Qisheng Li, {\normalfont and} Tim Althoff \\
}
\affiliations {
  Paul G. Allen School of Computer Science \& Engineering \\
  University of Washington, Seattle, WA \\
  \{dwadden, taugust, liqs, althoff\}@cs.washington.edu
}

\begin{document}

\maketitle



\begin{abstract}

  Many people struggling with mental health issues are unable to access adequate care due to high costs and a shortage of mental health professionals, leading to a global mental health crisis. Online mental health communities can help mitigate this crisis by offering a scalable, easily accessible alternative to in-person sessions with therapists or support groups. However, people seeking emotional or psychological support online may be especially vulnerable to the kinds of antisocial behavior that sometimes occur in online discussions. Moderation can improve online discourse quality, but we lack an understanding of its effects on online mental health conversations.
  In this work, we leveraged a natural experiment, occurring across 200,000 messages from 7,000 online mental health conversations, to evaluate the effects of moderation on online mental health discussions. 
  We found that participation in group mental health discussions led to improvements in psychological perspective, and that these improvements were larger in moderated conversations. The presence of a moderator increased user engagement, encouraged users to discuss negative emotions more candidly, and dramatically reduced bad behavior among chat participants. Moderation also encouraged stronger linguistic coordination, which is indicative of trust building. In addition, moderators who remained active in conversations were especially successful in keeping conversations on topic. Our findings suggest that moderation can serve as a valuable tool to improve the efficacy and safety of online mental health conversations. Based on these findings, we discuss implications and trade-offs involved in designing effective online spaces for mental health support.

\end{abstract}



\section{Introduction}

Over 400 million people globally struggle with mental health challenges, with approximately 300 million experiencing depression~\cite{WHOMentalHealth}. Depression leads to economic costs totalling more than \$100 billion annually in the United States alone~\cite{Twenge2019AgePA}. Rates of serious psychological distress -- including suicidal ideation and suicide attempts -- have increased 71\% in adolescents and young adults since 2005~\cite{Twenge2019AgePA}. Although psychotherapy and social support can be effective treatments~\cite{wampold2015great,WHODepression}, vulnerable individuals often have limited access to therapy and counseling~\cite{Bose2018}.

Instead, more and more people are turning to online mental health communities to express emotions, share stigmatized experiences, and receive helpful information~\cite{eysenbach2004health}.
These communities offer an accessible way for users to connect to a large network of peers experiencing similar challenges. Participants unable to access other treatment options can find social support and relief through these conversations~\cite{de2014mental,sharma2018mental,naslund_aschbrenner_marsch_bartels_2016}.
Recently, social support networks have begun to offer a more personalized experience by matching people sharing similar struggles in live, private conversations for support~\cite{althoff2016large}.

While online mental health communities can provide a valuable setting for giving and receiving support, the quality of support provided by peers is less well-characterized. Can conversation participants temporarily assume the role of a psychological counselor to assist those in serious distress? In addition, the often unrestricted and anonymous environment of online discussions can become a platform for anti-social behavior, such as online abuse or harassment~\cite{cheng2015antisocial,zhang2018conversations}. Are these concerns relevant in the setting of an app designed expressly for mental health discussion? Perhaps users of this platform are more thoughtful and considerate than the average forum participant. On the other hand, if bad behavior is an issue, moderation has been shown to be effective tool to combat undesirable behavior in online discussions~\cite{seering2019moderator,matias2019preventing,lampe2014crowdsourcing,Seo2007}. But little is known about the effectiveness of moderation in the context of mental health applications. Do moderators need to be highly \emph{involved} to keep users safe? Or does simply the knowledge that a moderator is \emph{present} influence behavior without active intervention? Furthermore, what roles do moderators assume in mental health discussions? Are they mostly discipline-keepers, or do they also act as counselors and facilitators?

In this work, we investigated how moderation affected online mental health conversations by identifying a natural experiment~\cite{DiNardo2016} occurring when the developers of an online platform hosting unmoderated mental health conversations discontinued the application, replacing it with a new platform in which all conversations were supervised by a moderator, while user population and application design remained largely identical. Together, these two platforms generated an extensive chat history of roughly 200,000 messages, constituting the largest dataset of its kind.
By comparing the linguistic attributes of the unmoderated and moderated conversations, we were able to extract data-driven insights about the effects of moderation on the civility, supportiveness, and outcomes of mental health discussions.  To ensure the validity of these findings, we performed a number of additional experiments to confirm that our results were attributable to the switch in moderation status, and were not artifacts caused by other factors such as the time frame of data collection or the number of conversation participants. In addition, we took advantage of naturally occurring variation in the degree of moderator activity -- i.e. how frequently moderators post in conversations -- to disentangle whether effects were simply due to moderator \textit{presence} or truly their active \textit{involvement.} Finally, we compared the behavior of unmoderated users to the behavior of moderators to determine whether users ever engage in counseling-like behavior to support their peers in settings without a moderator present.

Our findings include:

\begin{enumerate}[itemsep=0.01pt]
\item Users were more engaged in the conversation when a moderator was present, writing twice as many messages on average (Section \ref{sec:engagement}).

\item Moderators and users exhibited distinctive word usage patterns indicative of their roles in conversations. Moderators often acted as mental health counselors, emphasizing words involving perception and understanding others. Most users focused on explaining their concerns and struggles. They tended to disclose negative emotions more openly when a moderator was present. Interestingly, in the absence of a moderator, some users assumed a counseling-like role for other users (Section \ref{sec:word_usage}).

\item Conversations were much more likely to remain civil and free of toxic or harmful language when a moderator was present (Section \ref{sec:civility}).

\item Users coordinated more to one another in the presence of a moderator, suggesting stronger group cohesion and social support (Section \ref{sec:accommodation}).

\item Participants experienced positive perspective changes as measured by several psycholinguistic indicators. These improvements were larger on average in moderated conversations (Section \ref{sec:perspective}).

\item \emph{Actively} moderated conversations tended to stay more on-topic than passively moderated ones (Section~\ref{sec:relatedness}).

\end{enumerate}

In summary, moderation may positively impact online mental health conversations by encouraging more civil, on-topic conversations with higher user engagement. Based on these findings, we discuss design implications for online mental health platforms and how they may best utilize moderation to keep psychological support group conversations civil and beneficial for everyone (Section \ref{sec:discussion}).



\section{Related Work}

Our study is motivated by work analyzing the effect of moderation on online discourse (Section \ref{sec:moderation_effects}), examining the risks and benefits of online mental health communities (Section \ref{sec:online_health_communities}), and understanding digital mental health interventions (Section \ref{sec:digital_interventions}).

\subsection{The Effect of Moderation on Discourse} \label{sec:moderation_effects}

Civility and politeness are important elements of online communities~\cite{danescu2013computational,burke2008mind}, and can cause derailment of otherwise healthy discussions~\cite{zhang2018conversations}. Moderation as a way of encouraging civility is ubiquitous in online discussions, where anony\-mi\-ty can invite harassment and anti-social behavior~\cite{kraut2012building}. Moderators use a number of tools to maintain productive discussions, such as example-setting, posting rules on discussion threads, and restrictive bans when necessary~\cite{seering2019moderator,matias2019preventing}. ~\citet{lampe2014crowdsourcing} showed that moderation reduces uncivil and inflammatory rhetoric and encourages civil conversations in online discussions. In the context of classrooms,~\citet{Seo2007} found that students engage more actively and stay more on topic in peer-moderated online discussions than in unmoderated forums. ~\citet{seering2019moderator} interviewed 56 moderators across 3 major online platforms, finding that moderator involvement in a community can range from very active to only intervening when a serious transgression occurred. This also highlights different roles a moderator can fill: either as a facilitator, supporting conversations proactively, or solely keeping the peace of the online space~\cite{seering2019moderator}. In this work, we conduct the first large-scale analysis on the role of moderators in online mental health conversations and examine the effect of moderator \emph{activity} (in addition to presence) on discourse quality. In addition, this work represents the first study of the effects of moderation in the online mental health setting.

\subsection{Online Mental Health Communities} \label{sec:online_health_communities}

Online mental health communities are a valuable resource for peer-to-peer support~\cite{eysenbach2004health} due to their ease of accessibility, low cost, and the ability to remain anonymous. Work has shown that the ability to remain anonymous in computer mediated communication can increase self-disclosure~\cite{joinson2001self}, and~\citet{andalibi2016understanding} showed this in mental health communities by exploring how people shared stigmatized experiences in online mental health communities on Reddit. They found that many users use throwaway accounts (i.e., accounts with no personal information) for sharing these experiences as a way of maintaining anonymity. Throwaway accounts also have been reported to share content with increased negativity and self-focus, and lower self-esteem, supporting their use for self-disclosure~\cite{pavalanathan2015identity}.

Many factors play a role in a user self-disclosing, including a desire to manage impressions, online group size, and tie strength~\cite{10.1145/2818048.2820010}.~\citet{newman2011s} interviewed people to see how they share mental health information online, and identified two competing tensions of wanting to share information concerning a health issue and managing their own self-presentation. Work has also shown that mental health disclosure online can lead to positive outcomes such as emotional and informational support~\cite{de2014mental,sharma2018mental}, and positive cognitive change~\cite{pruksachatkun2019moments}.

~\citet{Webb2008ProvidingOS} and~\citet{Lederman2014ModeratedOS} describe the process of creating online mental health forums for adolescents with general mental health issues and psychosis, respectively. They identified roles for moderators including fostering a positive atmosphere, reporting crisis posts, setting boundaries, and encouraging users to practice cognitive and behavioral self-care skills.~\citet{zhang2020balancing} studied how mental health counselors in a crisis text line balanced responding empathetically and moving a conversation towards a resolution. Previous works have generally analyzed interactions occurring on public message boards or in crisis counseling. In this work, we instead leverage our naturally occurring dataset to understand how moderation affects private mental health discussions.

\subsection{Digital Mental Health Interventions} \label{sec:digital_interventions}

Researchers have explored more active interventions for supporting mental health.~\citet{shing-etal-2018-expert} developed an automated suicide risk assessment model based on Reddit posts rated by clinicians.~\citet{de2013predicting} built a classifier to predict the onset of depression from a users' social media posts.~\citet{Saha_De_Choudhury_2019} used social media data to evaluate the effects of psychiatric drugs, showing the feasibility of augmenting clinical studies with large-scale social media analyses of drugs' effects. Other recent work has characterized the ethical tensions of automated mental health interventions~\cite{chancellor2019taxonomy}, identifying issues such as construct validity and bias, and data privacy.



\section{Dataset}

We describe the dynamics of the online mental health mobile application, the chat log dataset studied in this work, and basic preprocessing steps used to filter out low-quality data.


\subsection{Dataset Description} \label{sec:dataset_description}

Our data consist of two sets of conversation logs from two mobile health platforms created by the same developers and sharing a similar user population and UI, but differing with respect to moderation status. Approval to analyze the two datasets was obtained from the Institutional Review Board at our institution.

\begin{figure}[t]
  \includegraphics[width=\columnwidth]{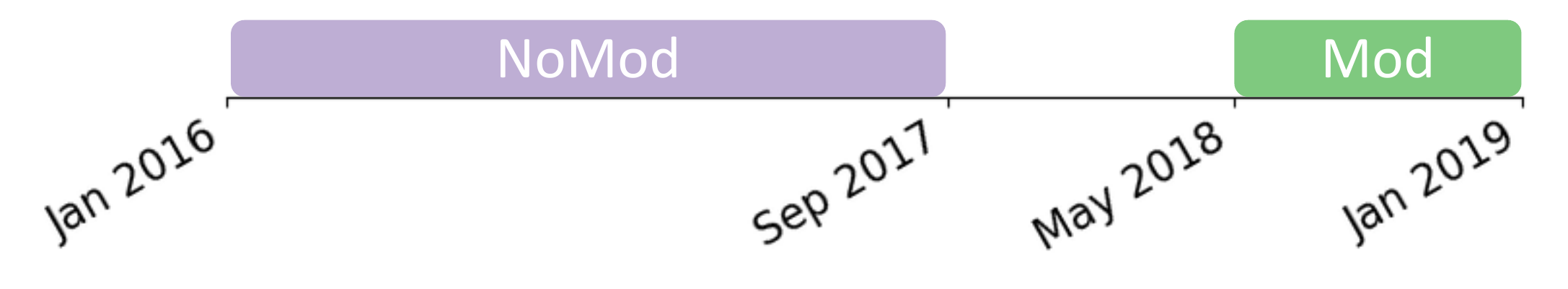}
  \caption{Data collection timeline.}
  \label{fig:timeline_simple}
\end{figure}

\begin{figure}[t!]
  \centering
  \includegraphics[width=\columnwidth]{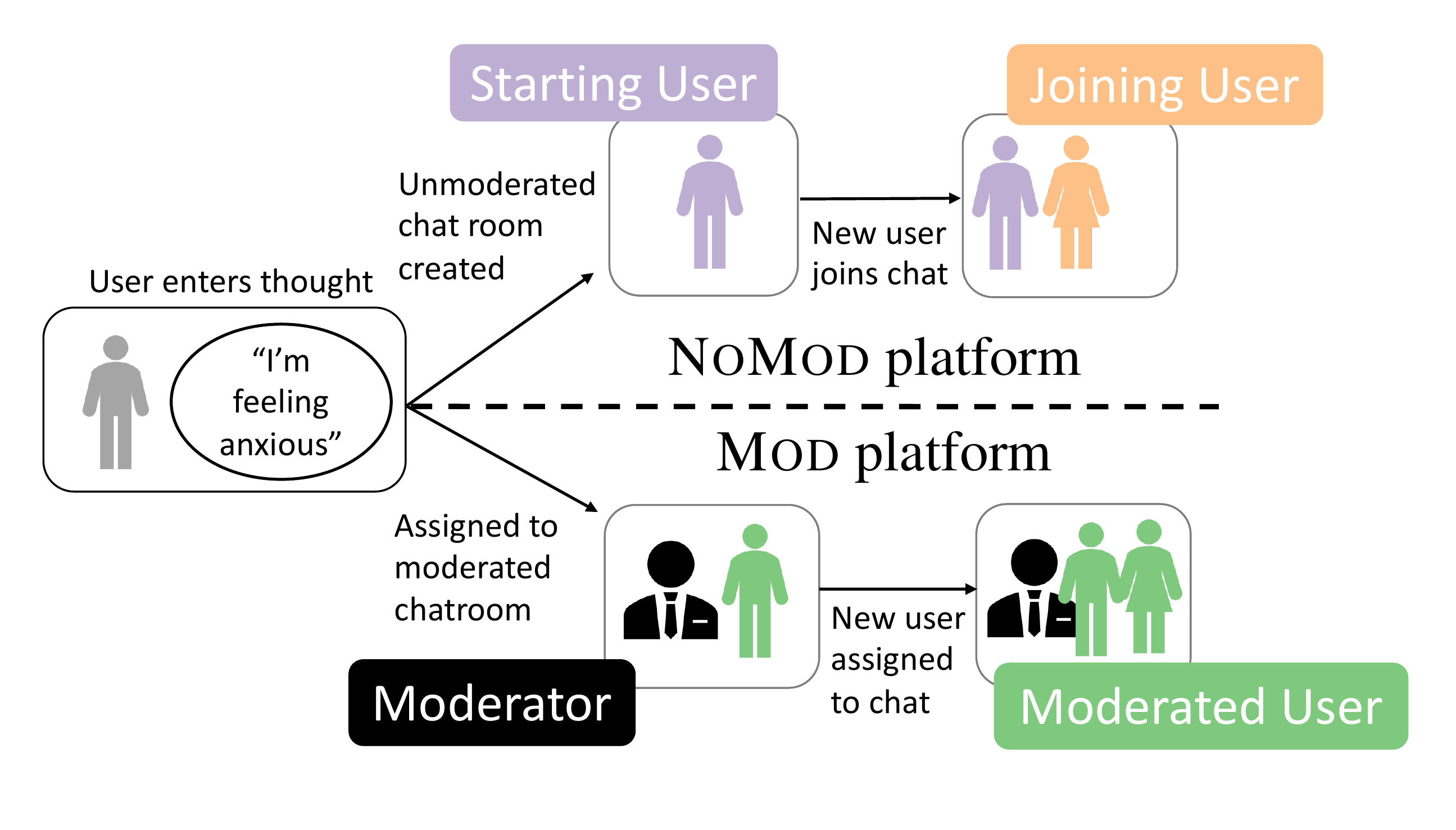}
  \caption{The life cycle of a conversation in \nmod\ (top) and \ymod\ (bottom). Participants are colored based on their conversation roles. In \nmod, participants are either \emph{starting users} (purple), or \emph{joining users} (orange). In \ymod, they are either \emph{moderators} (black) or \emph{moderated users} (green).}
  \label{fig:convo_illustration}

\end{figure}

Moderators were not present for conversations taking place on the first platform -- referred to as \nmod\ -- but were present for the conversations on the subsequent \ymod\ platform. The data collection timeline is shown in Figure \ref{fig:timeline_simple}. The life cycle of conversations in \nmod\ and \ymod\ is similar and is shown in Figure \ref{fig:convo_illustration}.

In \nmod, a \emph{starting user} wrote a post on a public topic page, creating a chat room. Other \emph{joining users} could view the subject line of the post and were free to join the chat room to discuss the content of the post. The conversation ended when all users exited the chat. Starting and joining users are referred to collectively as \emph{unmoderated users}.

In \ymod, conversations took place in persistent ``chat rooms'' presided over by a single moderator. Moderators were undergraduate or graduate students pursuing degrees in psychology, who had completed training by the platform on peer support facilitation. They were paid \$15 / hour. Upon opening the app, \emph{moderated users} were asked to write a sentence describing the issue they were struggling with. They were assigned to a chat room discussing topics relevant to their issue (e.g. depression, anxiety, relationship problems). The assignment was performed automatically by the app using natural language processing techniques. Depending on the number of users on the app and the similarity of their issues, a single chat room could host multiple users with related concerns, or host a one-on-one conversation between the room's moderator and a user. All chat participants were aware of the presence and identity (i.e. chat username) of the moderator.

In both \nmod\ and \ymod, users were assigned to conversations based on their interests and concerns. In \nmod\ this assignment was performed by the users themselves, while in \ymod\ it was performed by an algorithm. The app designers report that \ymod\ users were generally happy with their automatically-generated room assignments. Thus we have no reason to believe that this minor difference in assignment mechanism affected any of the outcomes studied.


\subsection{Data Preprocessing}

Our unit of analysis in this work is a single conversation. Therefore, we segmented the long-running chat room logs in \ymod\ into self-contained conversations by assuming that an interval of more than 15 minutes without a message indicated the end of a conversation\footnote{This cutoff was chosen in consultation with the app creators and validated with manual inspection.}.

We applied filters to remove very short, low-quality conversations. For \nmod, we kept all conversations with at least 2 participants and at least 10 messages long. For \ymod, we kept all conversations with at least 1 user, one moderator, and 10 messages. For both datasets, we kept conversations with a median message length of at least 5 tokens. 34\% of all messages in \nmod\ passed our filters, compared to 66\% for \ymod. All experiments were performed on the filtered data, for which Table~\ref{tbl:dataset_summary} provides summary statistics. The \nmod\ dataset is roughly 13 times larger than \ymod, and as a result the confidence intervals on \ymod\ estimates in the following sections are larger. While the smaller number of \ymod\ samples increases the \emph{variance} in the parameter estimates for this dataset, there is no reason to believe it would \emph{bias} our results in a particular direction. We address sources of potential bias in the next subsection.


\subsection{Considering Potential Threats to Validity} \label{sec:validity_threats}
We considered multiple potential threats to the validity of our reported findings.
If conditions are not randomized to participants, it is possible that any observed effects may be due to factors unrelated to the conditions of interest. We identified four potential factors and performed additional experiments to ensure that our findings were robust to them.

\begin{table}[t]
  \centering
  \begin{filecontents*}{tbl/dataset_summary_updated.tsv}
Index	NoMod	Mod
N conversations	7,079	317
N users	7,344	558
N messages	183,570	14,985
Median messages per chat	15	30
Median tokens per message	8	8
\end{filecontents*}
\csvreader[autobooktabular, separator=tab, table head=\toprule\ & \nmod & \ymod \\ \midrule ] {tbl/dataset_summary_updated.tsv}{Index=\data, NoMod=\nomodcol, Mod=\modcol}{\data & \nomodcol & \modcol}
  \caption{Summary statistics for filtered \nmod\ and \ymod.}
  \label{tbl:dataset_summary}
\end{table}

\xhdr{Time frame} As shown in Figure \ref{fig:timeline_simple}, the \nmod\ period ran for nearly 2 years, while \ymod\ ran for roughly 9 months. We performed experiments on appropriately-chosen subsets of \nmod\ to confirm that our findings were not driven by seasonality effects or differences in the lengths of the \nmod\ and \ymod\ data collection time frames. All experiments gave qualitatively similar results matching those reported in the main paper, which use all the available data (see Appendix \ref{appx:time_frame} for details).

\xhdr{Shifts in discussion topic} To establish that our findings were not an artifact of shifts in discussion topics between \nmod\ and \ymod, we confirmed with the app creators that there were no deliberate changes in the topics discussed on the app. Most discussions centered around everyday challenges related to, for instance, anxiety or depression related to work or relationships. As an additional check, we fit an LDA topic model and confirmed that the topic distributions were quite similar for \ymod\ and \nmod\ (see Appendix \ref{appx:topics}). This suggests that the topics of discussion were comparable between the two conditions.

\xhdr{Number of conversation participants} Conversation size varied in both \nmod\ and \ymod, and tended to be smaller for \ymod\ (Section \ref{sec:engagement}). To account for this, we initially stratified all analyses by the number of conversation participants. The results were qualitatively similar across different numbers of participants (see Appendix \ref{appx:convo_size} for an example). For ease of presentation, we show the unstratified results.

\xhdr{First-time vs. repeat users} We confirmed in discussions with the app creators that, due to the 8-month gap between the end of the \nmod\ version of the app and the launch of \ymod\ version, most \ymod\ chat participants were new users unfamiliar with the older \nmod\ version of the app\footnote{For the launch of \ymod, the app creators sent a promotional email to users of the old \nmod\ platform; fewer than 10 recipients tried the new platform.}. Thus, our findings are unlikely to be influenced by differences in behavior between new versus returning users.


\subsection{Data Anonymization}
Extra precautions were taken to anonymize all conversations, posts, and discussion topics. Following best practices dealing with stories around abuse~\cite{matthews2017stories}, we anonymized all presented quotes and conversations by removing or deliberately changing any identifying information, including generalizing specific mentions of places or people. We also added and removed filler words or rephrased content with less unique word choices or phrases. This process was repeated independently by two of the authors.
Due to the sensitive nature of the data, we are unable to provide screenshots of the mobile app as this could compromise the anonymity of participants. Instead, we are sharing several anonymized conversations based on the process described above (see Section~\ref{sec:word_usage}).



\section{Effects of Moderation on Conversation Dynamics and Outcomes} \label{sec:analysis}

We study the word usage patterns of moderators and users (Section \ref{sec:word_usage}), and the effects of moderation on user engagement (Section \ref{sec:engagement}) and civility (Section \ref{sec:civility}). We then explore how moderation promoted linguistic coordination, which is suggestive of supportiveness and group cohesion (Section \ref{sec:accommodation}), and facilitated positive changes in user perspective (Section \ref{sec:perspective}). Finally, we examine whether moderation helped keep conversations more on topic (Section \ref{sec:relatedness}).

We initially performed all analyses stratified into three groups by moderator \emph{activity}, measured by the fraction of messages sent by the moderator. We present stratified results in Section \ref{sec:relatedness}. For all other experiments, moderator activity level did not affect the outcome and we present unstratified results. See Appendix \ref{appx:moderator_activity} for statistics on moderator activity.


\subsection{User Engagement and Participation} \label{sec:engagement}


\begin{table}[t]
  \centering
  \begin{tabularx}{8.3cm}{l l l r r r}
    \toprule
    & Participant & Dataset & Q25 & Q50  & Q75  \\
    \midrule
    & User        & \nmod   & 3.4 & 5.3  & 8.5  \\
    Messages & User        & \ymod   & 5.3 & 10.0 & 19.0 \\
    & Moderator   & \ymod   & 8.0 & 13.0 & 29.0 \\
    \midrule
    & User        & \nmod   & 4.0 & 8.0  & 17.0 \\
    Tokens   & User        & \ymod   & 3.0 & 7.0  & 14.0 \\
    & Moderator   & \ymod   & 5.0 & 9.0  & 15.0 \\
    \bottomrule
  \end{tabularx}

  \caption{Messages per chat participant, and tokens per message.}
  \label{tbl:convo_length_stats}
\end{table}


We examined the engagement of users in moderated and unmoderated discussion. As shown in Table \ref{tbl:convo_length_stats}, users sent roughly twice as many messages per conversation in moderated discussions, indicating greater user engagement. Message length was similar across datasets.

70\% of conversations in \nmod\ had more than two participants, compared to 40\% in \ymod. One explanation for this change might be that users lost interest in supporting their peers when a moderator was present, instead waiting for a 1-on-1 conversation. On the other hand, the difference could be a simple consequence of the constant availability of the moderators. We return to this issue in Section \ref{sec:accommodation}.


\subsection{Word Usage of Moderators and Users}\label{sec:word_usage}

We expected that moderators and users might play different roles in conversations, with moderators counseling and guiding users through their emotional difficulties, as has been shown in prior work~\cite{zhang2020balancing}. To evaluate this possibility, we examined whether these different conjectured roles manifested themselves through distinctive word usage patterns. Additionally, we examined whether some unmoderated users might behave in a counseling role to help peers in need of support.

\xhdr{Methods} We categorized the chat messages into four groups based on the roles described in Section \ref{sec:dataset_description}: starting users, joining users, moderated users, and moderators.

\begin{figure}[t]
  \centering
  \includegraphics[width=\columnwidth]{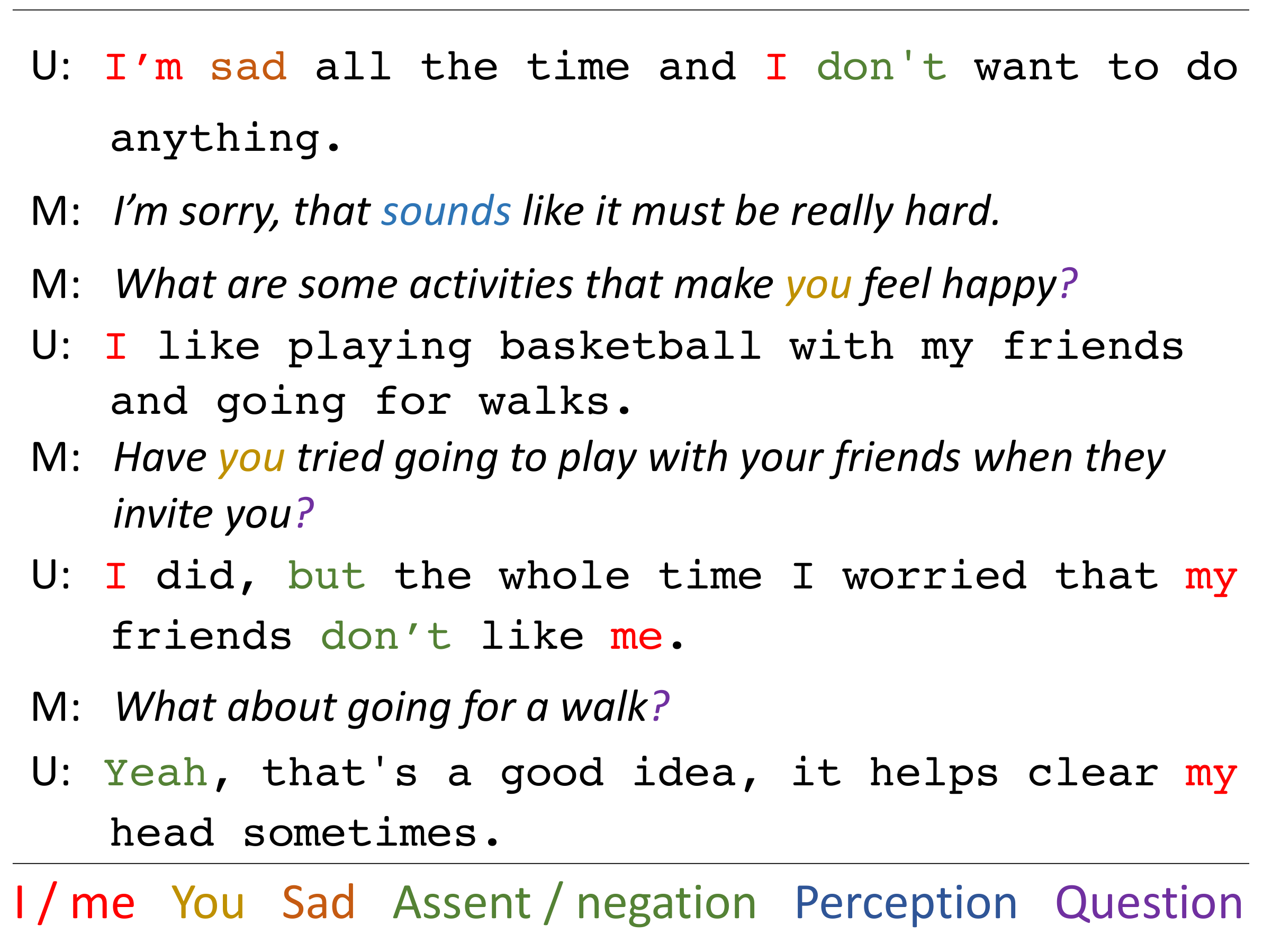}
  \caption{A conversation excerpt, with content altered to preserve anonymity.}
  \label{fig:example_convo}
\end{figure}

We used the LIWC lexicon~\cite{tauczik2010psychological} to quantify differences in word usage among our four groups. LIWC defines 76 psycholinguistic categories -- for instance ``Sad'' -- and provides a list of words associated with each category. Figure \ref{fig:example_convo} shows an example conversation annotated with LIWC categories. In Figure \ref{fig:example_convo} and throughout, we denote quotations from moderators with \moderator{italicized text} and quotations from users with \user{typewriter text}. For each category, we computed relative changes in word usage for each group of users, compared to moderators. Word usage was measured by computing the fraction of words in each utterance belonging to each LIWC category (see Appendix \ref{appx:math}).

\xhdr{Results} Figure~\ref{fig:liwc_moderators_vs_users} shows usage of each LIWC category by each of the three user groups, relative to usage by moderators. Overall, we found that moderators and users showed speech patterns consistent with their conversation roles. Joining users also exhibited some counseling behavior similar to moderators. Validity checks confirmed that these results were robust to the factors described in Section \ref{sec:validity_threats} (see Appendix \ref{appx:validity_threats} for some examples).

\begin{figure*}[t!]
    \centering
    \includegraphics[width=\textwidth]{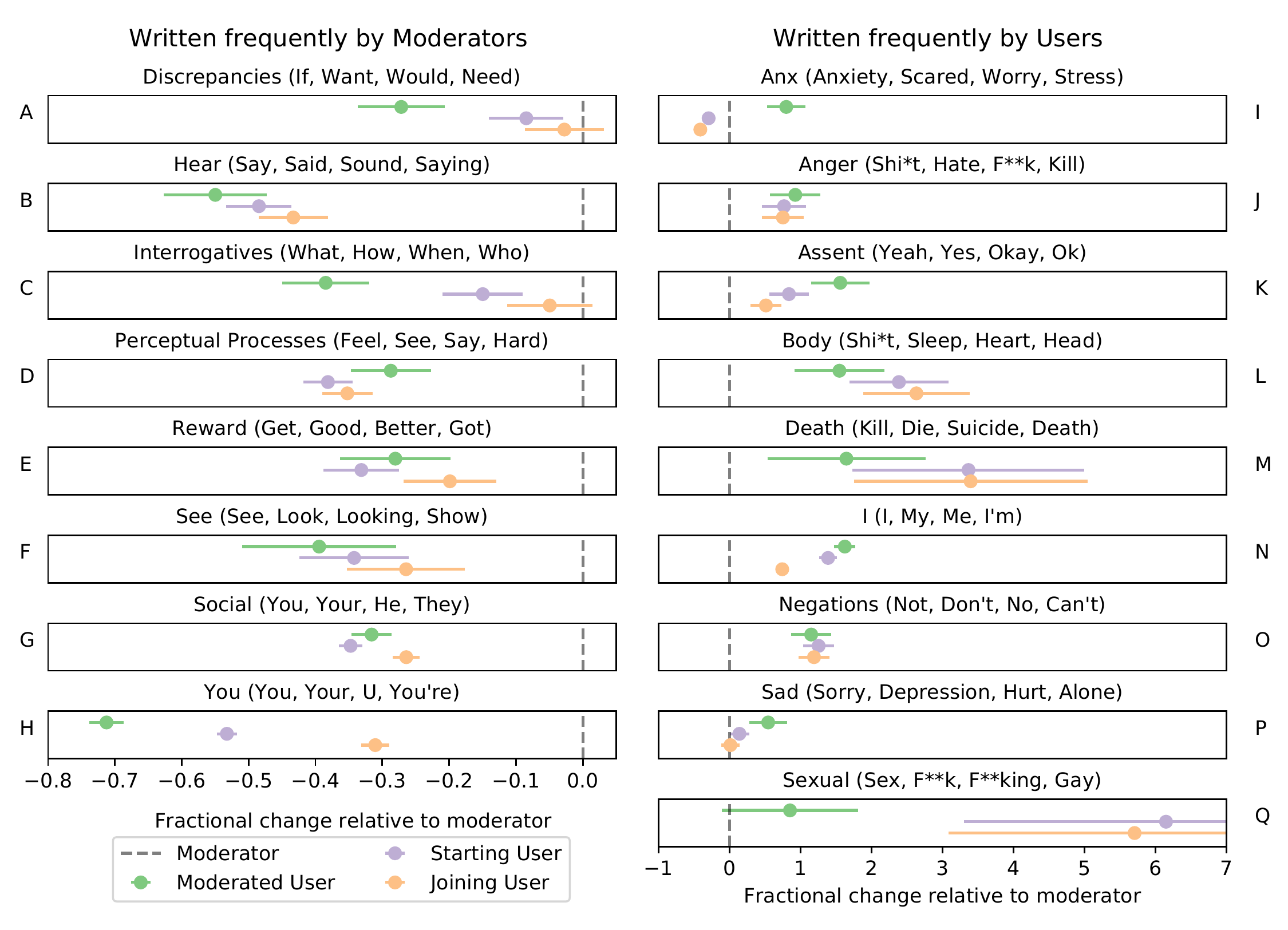}
    \caption{Word usage of different user groups. The x-axis shows the change in word usage of each LIWC category relative to moderator usage. For instance, a value of 1.2 indicates a 120\% increase in word usage. Moderator usage is represented by dashed vertical lines. In this figure and throughout the paper, error bars are bootstrapped 95\% confidence intervals. Any time the error bars between two groups do not overlap, the word usage between the groups is statistically significant at the 5\% level.}
    \label{fig:liwc_moderators_vs_users}
\end{figure*}

\xhdr{Users employed more self-focused language} Moderators expressed their interest in the well-being of users by making heavy use of language related to perception (Panels B, D, F), such as ``\moderator{Oh, I see}'' or ``\moderator{I hear where you're coming from}''.

\begin{sloppypar}
In addition, moderators used the most second-person pronouns (Panels G, H), for instance ``\moderator{That must make you feel lousy}''), and the fewest first-person pronouns (Panel N). Moderated users employed the most first person pronouns and the fewest second-person, issuing statements like ``\user{Sometimes when I'm nervous, I'll procrastinate}'' or ``\user{I've always fought with my siblings}''. These word usage patterns indicate that conversations were focused on the issues and concerns of users. Interestingly, a similar division occurs in unmoderated conversations: the pronoun usage of \emph{joining} users was more similar to moderators, while the usage of \emph{starting} users was more similar to moderated users. This suggests that counseling-type behavior occurred even in unmoderated conversations -- albeit by untrained users.
\end{sloppypar}

\begin{sloppypar}
  \xhdr{Users experienced negative emotions} Moderated users expressed feelings of anxiety (Panel I; ``\user{I worry that I'm unlovable}'') and sadness (Panel P; ``\user{I've been hurt and betrayed over and over}'') more readily than both moderators and unmoderated users, suggesting that the presence of a moderator may have allowed conversation participants to speak more openly about difficult emotions.

All users uttered words related to anger (Panel J; ``I hate my job'') and death / suicide (Panel M) more frequently than moderators. Some death-related utterances were benign, like ``\user{My job is killing me right now}'', but others indicated serious distress, for instance ``\user{I came close to killing myself}'' or ``\user{I wish I could die}''.

In response to the range of negative emotions voiced by users, moderators employed positive, encouraging language (Panel E), for instance ``\moderator{Physical activities like going for a walk are a good distraction from negative thoughts}'', ``\moderator{It's good that you have friends and family in your corner}''.
\end{sloppypar}

\begin{sloppypar}
\xhdr{Word meaning was influenced by context} Body-related (Panel L) and sex-related (Panel Q) words were employed by users for different purposes. In some cases,  words related to body parts or physical processes were used literally to discuss physical discomfort or health challenges, such as ``\user{I can't sleep at night because I can't stop thinking}'' or ``\user{I get bad headaches when I'm stressed}''. Other times, users discussed body parts metaphorically to convey emotions: ``\user{My heart is so lonely}'' or ``\user{I get in my own head and second-guess myself}''. Bodily functions were also used as curse words: ``\user{After all the sh*t, I still forgive them}'' or ``\user{That's crap}''.

A similar phenomenon occurred for sex-related words. Some users wanted to ask advice or discuss their sex lives: ``\user{I just had sex for the first time and I'm not sure how I feel about it}'' or ``\user{It's hard to come out, because my family is religious}''. Others used words concerning sex or sexuality as insults: ``\user{f*ck you}'' or ``\user{That's gay}''.

Unmoderated users employed body-related words slightly more frequently than moderated users and uttered sex-related words much more frequently. Much of this difference was due to the higher rates of profanity in unmoderated discussions, as explored in Section \ref{sec:civility}.

\end{sloppypar}

\xhdr{Moderators and joining users often acted as counselors} When moderators and users took part in a one-on-one conversation, they frequently followed a pattern of asking questions, making suggestions, and providing feedback. First, the moderator used interrogatives (Panel C) to ask questions and understand the user's situation, and the user responded with clarification. Then, the moderator made suggestions, often employing a LIWC category known as \emph{discrepancies} (Panel A). Moderators utilized discrepancies to give advice or show interest, while allowing the user to disagree or express preference for an alternative. For instance a moderator might say, ``\moderator{I'm happy to listen if you want to share your thoughts}'' instead of ``\moderator{tell me what you're thinking}''. In response, users provided feedback in the form of either assent (Panel K), indicating that they agreed with the moderator's advice or assessment or negation (Panel O), indicating that the advice was not right for them. The conversation excerpt below depicts a moderator asking questions to understand a user's situation and offering advice and resources.

\begin{dialogue}
    \userq{I'm so annoyed. My boyfriend has been seeing other women, and when I confront him he lies.}

  \moderatorq{So you've tried to talk with him? Does he get mad when you bring it up?}

  \moderatorq{I would try to really communicate to him how hurtful it is to you. Does that make sense?}

  \userq{It's just so confusing. He can be so nice to me, but he gets mad whenever I ask him.}

  \moderatorq{I'm with you. I would ask, too!}

  \moderatorq{I can send you an article about healthy communication if you'd like?}

  \userq{Ya I would really appreciate that.}
\end{dialogue}

Just as joining users exhibited different pronoun usage compared to starting users, they also also made slightly greater use of discrepancies and interrogatives. The following excerpt shows a discussion between a user \textsc{U1} who started a chat with the subject ``\user{I want to hurt myself}'', and a joining user \textsc{U2} who saw the subject line and offered help, acting like an informal counselor. We show messages from \textsc{U2} in \textit{italics} for readability.

\begin{dialogue}
\uone{I want to hurt myself.}

\utwo{Please don't!}

\utwo{Do you want to tell me about it?}

\utwo{I also struggle with this. You need to find a way to focus on other things.}

\uone{I've tried. I wish I could follow that advice but I can't fight the urge.}

\utwo{Maybe you could listen to some happy music, or draw?}

\uone{Yes, you're right I should do that.}

\utwo{Have you told anyone how you feel?}

\uone{Well I have really great friends and family, but I don't want to burden them.}

\utwo{You should tell them how you feel. They care about you and they will want to help.}
\end{dialogue}


\subsection{Conversation Civility}~\label{sec:civility}

Due to the potential vulnerability of mental health conversation participants, ensuring that discussions remain free of offensive, profane, and toxic language is a top priority. While previous research has shown that moderation can reduce profanity in online forums~\cite{lampe2014crowdsourcing}, no prior work has examined whether moderation is necessary -- or effective -- in the setting of an application designed exclusively for mental health support.

\xhdr{Methods} We used Google's TensorFlow toxicity identifier\footnote{\url{https://github.com/tensorflow/tfjs-models/tree/master/toxicity}} to examine the effect of moderation on conversation civility. The model identifies seven types of violent and abusive language: (1) identity attacks, (2) insults, (3) obscenity, (4) sexually explicit content, (5) threats, (6) toxicity, and (7) severe toxicity. We ran this identifier on all messages sent by users in \ymod\ and \nmod\ (removing moderator messages)\footnote{To confirm that our findings were robust to the particular software toolkit used, we also made profanity predictions using the \texttt{profanity-check} Python library (\url{https://github.com/vzhou842/profanity-check}) and obtained similar results.}. For the unmoderated conversations, initial experiments did not reveal any differences in profanity between starting users and joining users; therefore, we analyzed all unmoderated messages together. Fortunately, there were no instances of severe toxicity in either dataset, so this category was not considered further.

\begin{figure}[t!]
    \centering
    \includegraphics[width=\columnwidth]{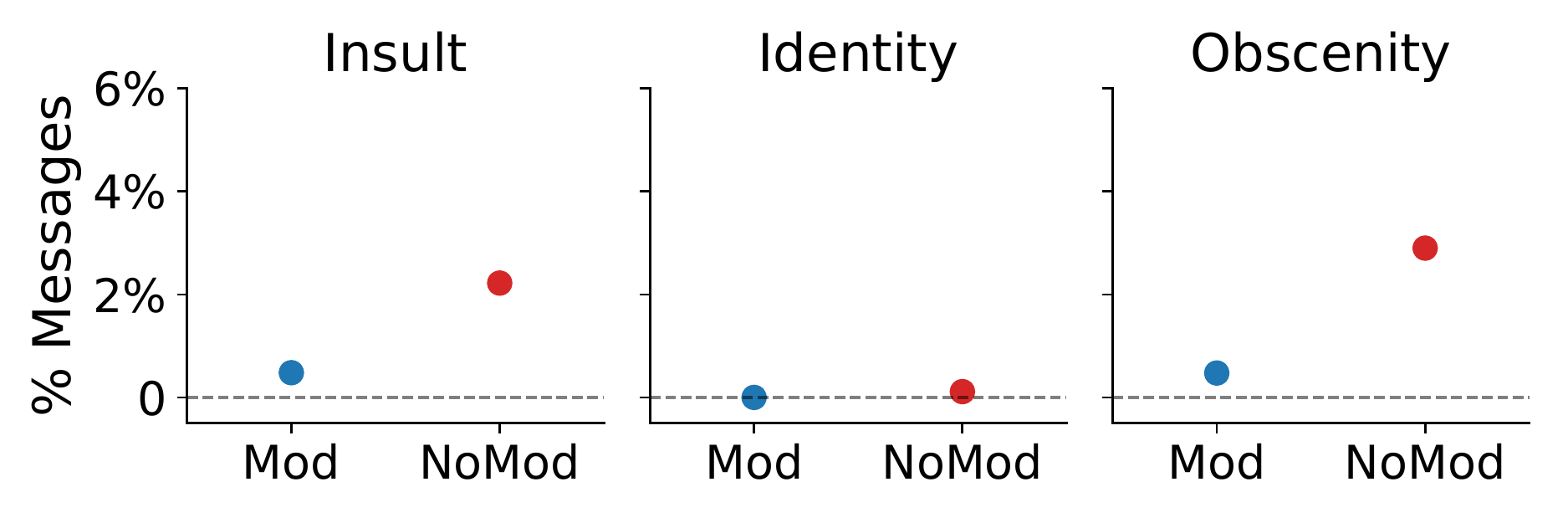}
    \includegraphics[width=\columnwidth]{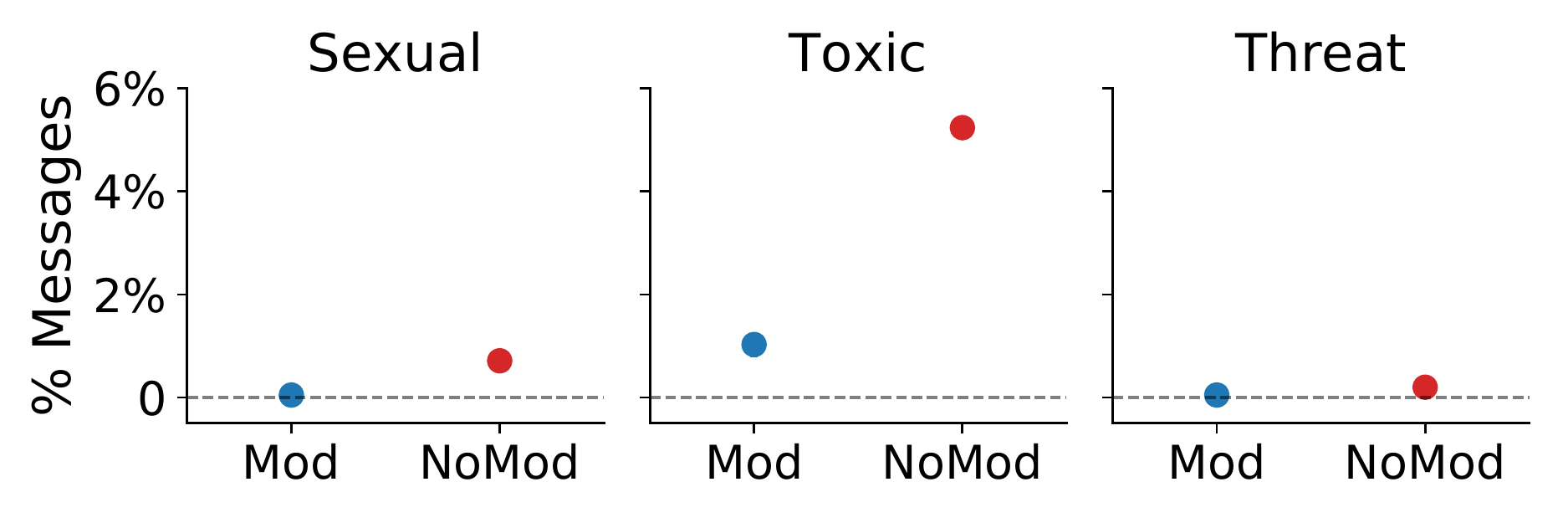}
  \caption{Percentage of user messages in each incivility category. Moderated conversations are almost totally free of toxic and harmful language. Error bars are computed but are too narrow to be visible.}
  \label{fig:civility}
\end{figure}

\xhdr{Results} Figure \ref{fig:civility} shows the percentage of \ymod\ and \nmod\ messages containing each category of offensive language. Moderation reduces all forms of incivility to 1\% of messages or lower, and reduces the occurrence of general toxic language five-fold -- from 5.2\% of messages to 1.0\%. These results suggest that moderation is, indeed, highly effective at combating toxicity and incivility in online mental health settings. Moderators seem to have accomplished this without coming across as disciplinarians and driving users away, as users tended to stay longer in moderated chats compared to unmoderated ones (Section \ref{sec:engagement}).

\subsection{Coordination and Trust-Building}~\label{sec:accommodation}

Given our results in Section \ref{sec:engagement}, we wondered whether users might engage less with their peers when a moderator was present, preferring 1-on-1 counseling conversations instead. To evaluate this possibility, we analyzed the degree to which users in \nmod\ and \ymod\ coordinated linguistically toward one another. Past work suggests that higher levels of coordination are indicative of trust-building~\cite{scissors2008linguistic} and lead to improved group performance~\cite{fusaroli2012coming} and social support~\cite{sharma2018mental,danescu2013no}.

\xhdr{Methods} We used the Cornell Convokit\footnote{http://convokit.cornell.edu/} to measure linguistic coordination among users in both datasets. Convokit measures linguistic coordination with words processed non-consciously by listeners and unrelated to topic, such as articles, auxiliary verbs, and conjunctions~\cite{danescu2012echoes}.

\begin{figure}[t!]
  \centering
  \includegraphics[width=\columnwidth]{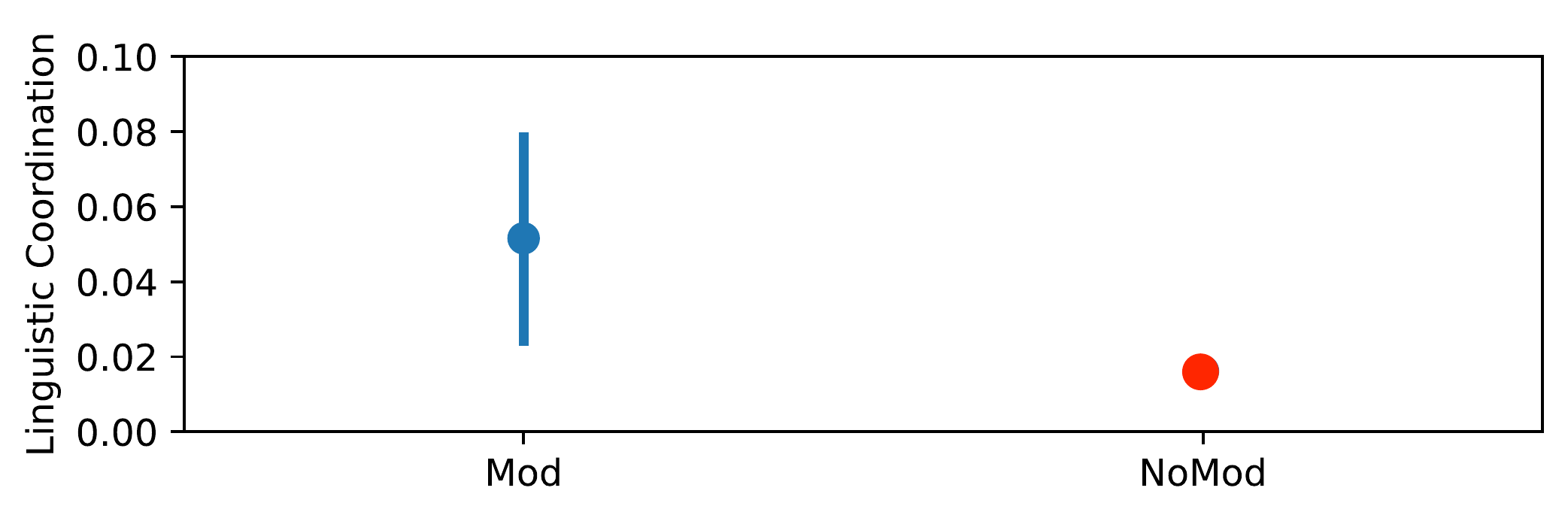}
  \caption{Linguistic coordination in \nmod\ and \ymod. On average peer users in \nmod\ coordinated significantly less to one another (mean=0.01, med=0, sd=0.07) than users in \ymod\ did to one another (mean=0.05, med=0.01, sd=0.13) ($U$ = 433419, $p<0.01$)}
  \label{fig:ling_acc}
\end{figure}

\xhdr{Results} Surprisingly, the results in Figure~\ref{fig:ling_acc} demonstrate that on average peer users in \ymod\ actually coordinated significantly \emph{more} toward one another than users in \nmod. The higher levels of coordination in \ymod\ conversations suggest that users did not lose interest in group discussion. Instead, moderation positively impacted conversations by encouraging coordination.

\subsection{Positive Change in Perspective}~\label{sec:perspective}

Previous work has demonstrated that 1-on-1 online counseling conversations can facilitate positive changes in psychological perspective~\cite{althoff2016large}, but it is not clear how a group setting could influence psychological perspectives differently. We explored whether the different user groups experienced perspective changes consistent with their roles from Section \ref{sec:word_usage}.

\xhdr{Methods} We leveraged the LIWC lexicons to compute three different measures of psychological perspective, following the approach in~\citet{althoff2016large}.

\begin{itemize}[itemsep=0.01pt]
\item \textbf{Time}: Psychological research has linked depression with excessive rumination about past events~\cite{Pyszczynski1987DepressionSA}, while recent work examining 1-on-1 online counseling conversations has found associations between greater future-orientation and more positive conversation outcomes~\cite{althoff2016large}. We computed the usage of LIWC future words as a fraction of all future and past-related words. Higher values indicate more focus on the future and less on the past, suggesting a more positive perspective.

\item \textbf{Self}: Individuals experiencing depression can become preoccupied with their own thoughts and have difficulty engaging with others~\cite{Pyszczynski1987DepressionSA,Pyszczynski1987SelfregulatoryPA}. This tendency can be expressed through heavy use of first-person pronouns. We measured self-focus by computing the fraction of pronouns used by conversation participants that were first-person, as opposed to second or third person. Lower self-focus suggests a more positive perspective.
\begin{sloppypar}
\item \textbf{Sentiment}: To measure sentiment, we computed the usage of LIWC words related to positive emotion (PosEmo), as a fraction of the total number of words related to any emotion -- positive or negative (NegEmo). Higher values suggest more positive sentiment.
\end{sloppypar}
\end{itemize}

\begin{figure}[t!]
  \begin{subfigure}[t]{\columnwidth}
    \centering
    \includegraphics[width=\columnwidth]{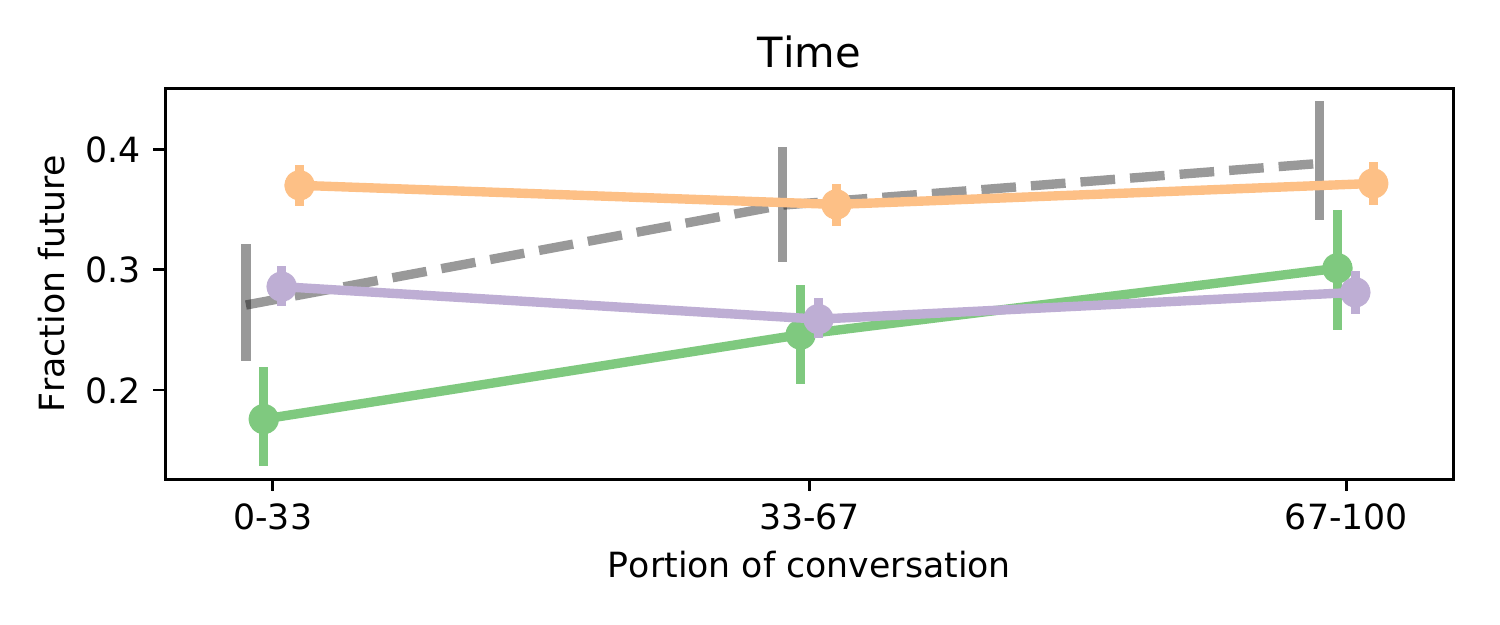}
    \subcaption{Of the three user groups, moderated users show the largest increase in future focus, rising by $0.13 \pm 0.06$ from the first third to the final third of each conversation. Reported uncertainties are 2 standard errors of the mean.}
  \end{subfigure}
  \begin{subfigure}[t]{\columnwidth}
    \centering
    \includegraphics[width=\columnwidth]{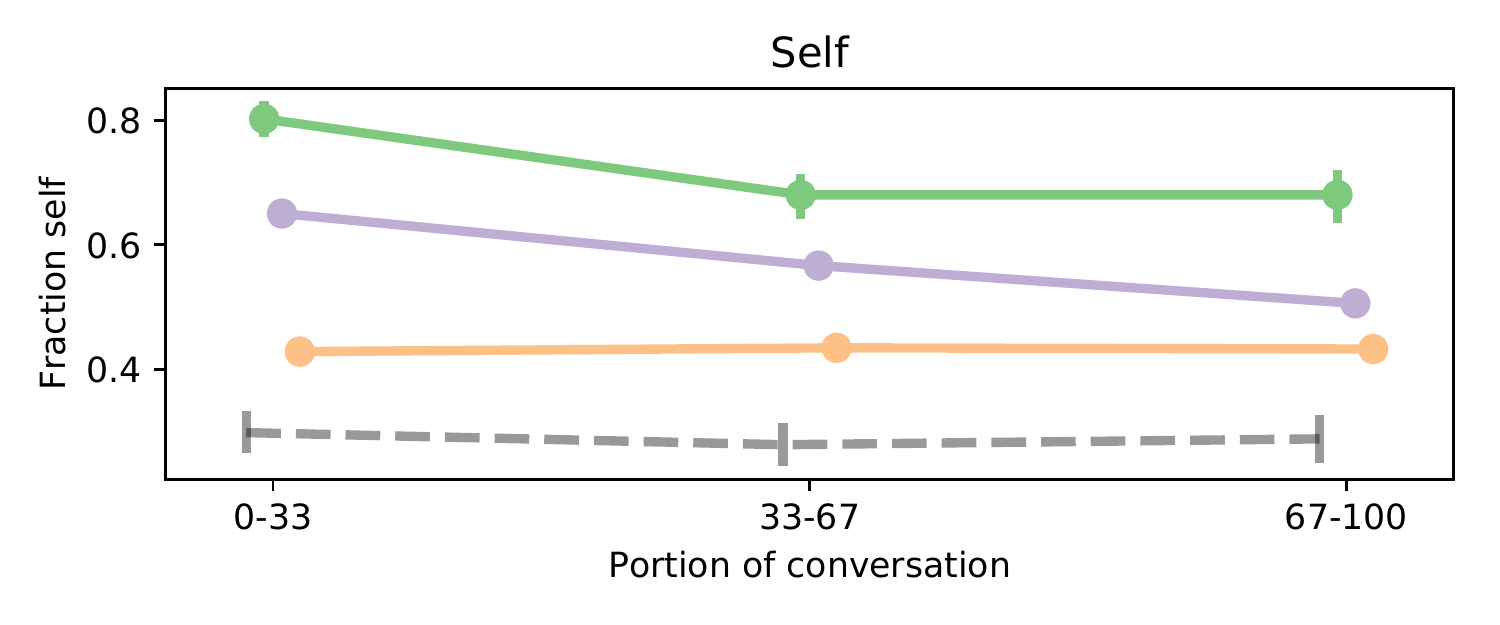}
    \subcaption{Moderated users and starting users show similar decreases in self-focus ($-0.12 \pm 0.04$ and $-0.14 \pm 0.02$, respectively).}
  \end{subfigure}
  \begin{subfigure}[t]{\columnwidth}
    \centering
    \includegraphics[width=\columnwidth]{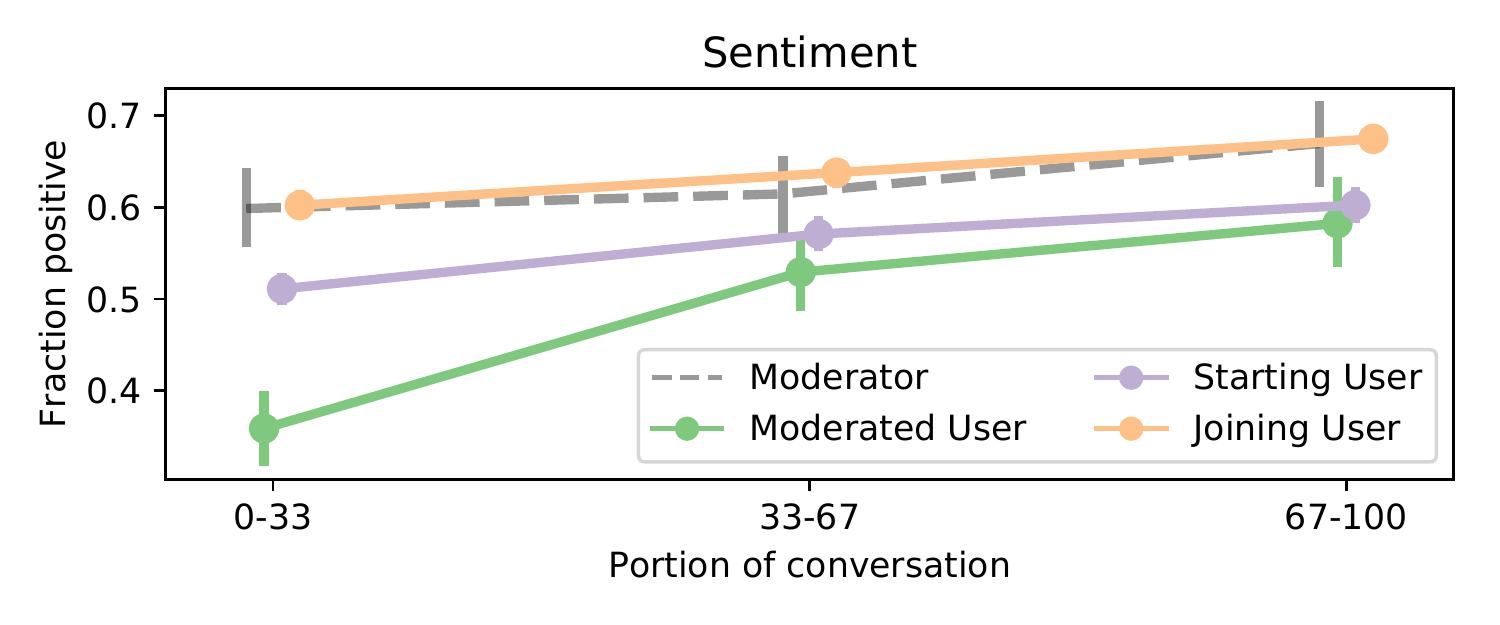}
    \subcaption{Moderated users show the largest sentiment increases ($0.22 \pm 0.05$), followed by starting users ($0.09 \pm 0.02$).}
    \label{fig:sentiment}
  \end{subfigure}
  \caption{Change in perspective over time. Moderators and joining users exhibit counseling-type behavior. Users experience significant improvement in perspective.}
  \label{fig:sentiment_over_time}
\end{figure}

To measure how user perspective changed over the course of a conversation, we divided each conversation into thirds, and computed the perspective measures for each third. To ensure that we had enough data for each time period, we restricted our analysis to conversation with at least 20 messages sent by users. This left 2,541 conversations in \nmod\ and 219 in \ymod. As in Section \ref{sec:word_usage}, we grouped messages into those sent by moderators, moderated users, starting users, and joining users. Moderator activity did not have a substantial effect on perspective, so all moderated users were analyzed together.

\xhdr{Results}  Figure \ref{fig:sentiment_over_time} reveals important differences in perspective trajectory for the three user groups\footnote{Validity checks confirmed that these differences were robust to variations in time frame and chat room size (Appendix \ref{appx:validity_threats}).}. Joining users behaved very similarly to moderators. They showed positive perspective early on and maintained it throughout their discussions, consistent with their hypothesized role as counselors for distressed conversation starters (Section \ref{sec:word_usage}). Moderated users showed the most negative perspective initially, consistent with the observation that moderated users express negative emotion more freely (Section \ref{sec:word_usage}). Fortunately, they also enjoyed the largest overall improvement in perspective, and by the end of discussion their levels of sentiment and future-focus were nearly as high as those of unmoderated users. Their self-focus also decreased, but remained higher than the self-focus of other users.

This is unsurprising in one sense, since moderators showed the lowest self-focus and likely continue eliciting first-person responses from users throughout their conversations. However, it is also surprising since moderated users coordinated more with one another than unmoderated users (see Section~\ref{sec:accommodation}), suggesting that although users might focus more on themselves they also were focused on engaging with other users in moderated conversations. Starting users had initial perspectives between those of joining users and moderated users. They showed improved sentiment and a greater decrease in self-focus than moderated users, consistent with their conversation role.


\subsection{Staying on Topic}~\label{sec:relatedness}

\begin{figure}[t]
  \centering
  \includegraphics[width=\columnwidth]{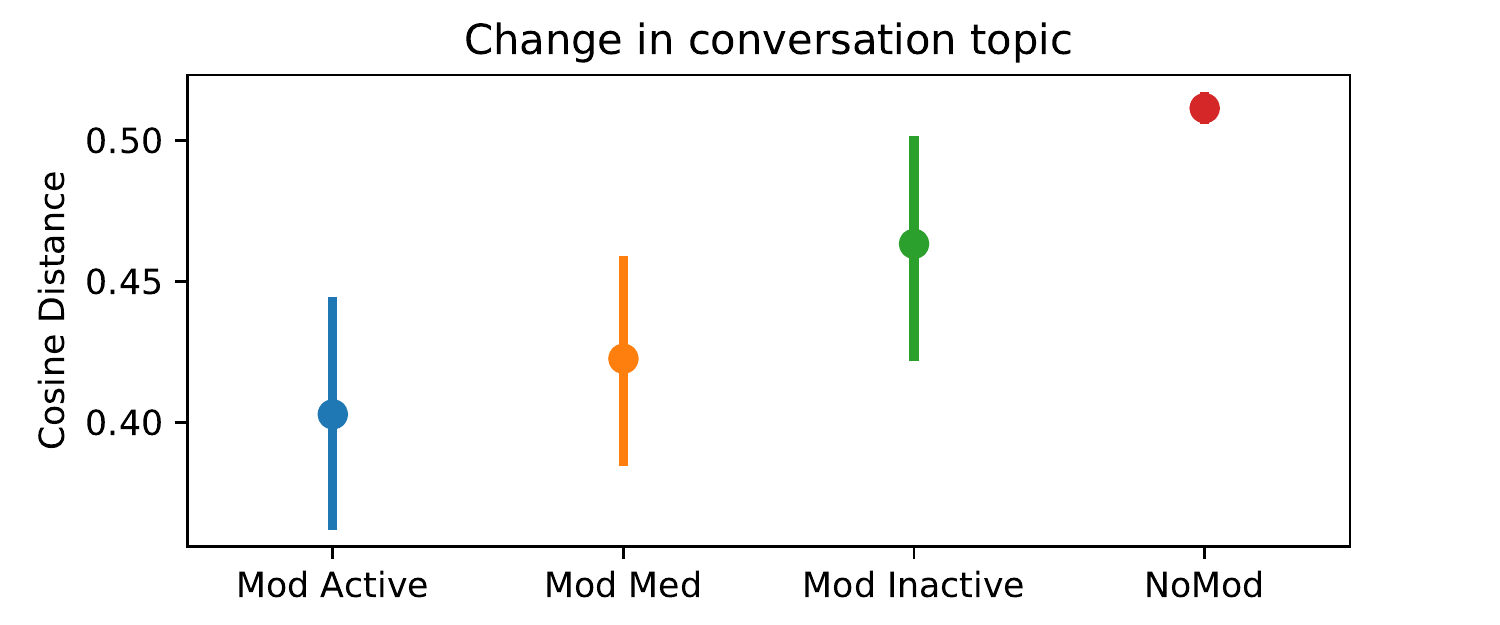}
  \caption{Mean cosine distances between first half and second half of conversations with heavy moderator activity (mean~=~0.40), medium activity (mean~=~0.42), light activity (mean~=~0.46) and with no moderation (mean = 0.51). The pairwise differences among the four groups of conversations are all statistically significant ($p < 0.05$) except for Mod Med compared to Mod Inactive ($p = 0.14$) or Mod Active ($p = 0.49$).}
  \label{fig:topic}
\end{figure}

In an online classroom setting, previous work has demonstrated that conversations remained more on topic when discussions were moderated~\cite{Seo2007}.  We explored whether this trend persisted in the more open-ended and personal conversations in our dataset, and further, whether \emph{active} moderation was necessary to achieve an effect.

\xhdr{Methods} Motivated by work in topic segmentation~\cite{Hearst1997TextTilingST}, we measured the degree to which a conversation remained on-topic by splitting each conversation in half and computing the cosine distance between bag-of-words representations of the two conversation halves. A lower distance indicates that the conversation stayed on a single topic.

In this experiment, we found that the outcome was influenced by the degree of moderator activity. We present results with the moderated data stratified into three equal-sized collections of conversations. \ymod\ Active contains the conversations with greatest moderator involvement, \ymod\ Inactive with the least involvement, and \ymod\ Med in between (see Appendix \ref{appx:moderator_activity} for more details).

\xhdr{Results} As shown in Figure~\ref{fig:topic}, we found that conversations stayed significantly more on topic in chats with a moderator present, in agreement with previous findings~\cite{Seo2007}. In addition, our results suggest that conversations with an active moderator may remain more on topic than those with an inactive moderator ($p < 0.05$).



\section{Discussion} \label{sec:discussion}

In this work, we performed the first large-scale quantitative analysis examining the effect of moderation on online mental health discussions, using the unmoderated and moderated conversation logs in our data set as control and treatment groups in a natural experiment. We found that:
\begin{itemize}[itemsep=0.01pt]
\item Moderation improved user engagement (Section \ref{sec:engagement}).
\item Moderators and users employed different language consistent with their conversation roles. In the absence of moderators, some users assumed a counseling role to support their peers (Section \ref{sec:word_usage}).
\item Moderation improved conversation civility (Section \ref{sec:civility}).
\item Moderation was associated with linguistic coordination, which was indicative of trust building and social support (Section \ref{sec:accommodation}).
\item Users in moderated conversations experienced larger positive perspective changes (Section \ref{sec:perspective}) and stayed more on topic (Section \ref{sec:relatedness}).
\end{itemize}

To support these conclusions, we performed a number of validity checks (Section \ref{sec:validity_threats} and Appendix \ref{appx:validity_threats}) and confirmed that:
\begin{itemize}[itemsep=0.01pt]
  \item The time frames over which the data were collected did not impact our findings.
  \item The topics discussed were highly similar in both \ymod\ and \nmod.
  \item While the number of conversation participants varied, it did not have an effect on our findings.
  \item Both \ymod\ and \nmod\ were predominantly composed of new users and therefore the presence of new vs. repeat users was not significantly different.
\end{itemize}

Below we discuss some implications of our findings.

\xhdrq{Should conversations be moderated}
The results in this work are consistent with findings on moderation in other web domains~\cite{matias2019preventing,lampe2014crowdsourcing}, and indicate that moderation may be an effective approach to ensure that participants stay safe while participating in potentially challenging discussions about mental health. Compared to unmoderated or inactively moderated conversations, our results also suggest that \textit{active} moderators may keep conversations more civil (Section \ref{sec:civility}) and on-topic (Section \ref{sec:relatedness}).

\xhdr{The role and limitations of peer support} Peer support can be helpful in encouraging personal connections and scaling social support~\cite{O'Leary:2018:WGB:3173574.3173905}. Indeed, we found ample evidence of \emph{user-as-counselor} behavior in our data (Sections \ref{sec:word_usage} and \ref{sec:perspective}). Users often responded to peers experiencing crises like suicidal ideation by seeking to understand and offer support, sending messages like ``\moderator{Please don't hurt yourself}'' or ``\moderator{Why would you want to kill yourself}?'' However, there were exceptions. Even the best-intentioned users are not trained to handle mental health crises and risk responding to vulnerable individuals in ways that could do unintended harm. In addition, we observed that some individuals experiencing mental health crises were angry and hostile, and used profane and abusive language (Sections \ref{sec:word_usage} and \ref{sec:civility}). We also uncovered some evidence of predatory behaviour in unmoderated chats (e.g., encouraging people to share social media accounts or pictures). Trained moderators are better suited than peers for dealing with acute and severe crises like suicidal ideation, hostility, and predatory behaviour, reducing the risk of harmful conversations.

\xhdrq{Who should be a moderator} Moderators for a mental health support site could vary in their training and expertise -- ranging from volunteers without psychological training to licensed social workers and psychologists. The undergraduate and graduate psychology students used as moderators in the application studied here occupied a middle ground.

The degree of moderator training required depends on the role that the moderator is expected to play in the conversations. On one extreme, a \emph{moderator-as-supervisor} could simply supervise the conversation, discouraging bad behavior (Section~\ref{sec:civility}) and stopping or appropriately escalating any discussions of self-harm. Based on our findings and previous work on moderation in public forums like Reddit~\cite{matias2019preventing,seering2019moderator}, it appears that \emph{any} moderator could fill this role.

On the other extreme, a \emph{moderator-as-counselor} would be expected to guide the conversation, identifying participants' issues and offering concrete suggestions (as in Section~\ref{sec:word_usage}), and keeping the conversation on topic (Section~\ref{sec:relatedness}). Through manual analysis of chat logs in our data set, we found that many moderators were in fact working in this capacity (as exemplified by the conversation excerpts from Section \ref{sec:word_usage}). Moderators serving as counselors would require appropriate training as determined by mental health professionals.

\xhdr{Limitations \& Future Work}
In this work we explored short-term mental health conversations taking place on one popular application platform, and found that moderation meaningfully improved discourse quality and improved the perspective of conversation participants.
As in any observational study, we cannot be totally certain that the assignment of users to the \nmod\ or \ymod\ group was independent of other factors affecting conversation outcome. However, our validity checks and conversations with the app creators provide good assurance that our findings are correctly attributed and are not spurious artifacts. In addition, we did not identify any obvious idiosyncrasies of the app which would prevent our findings from generalizing to other online mental health forums. This work represents a first attempt at understanding the effects of moderation in the online mental health setting, and we eagerly await the availability of additional datasets on which to validate the generalizability our findings.

In our analysis, we leveraged psycholinguistic tools in order to gain insights into the mental states of users. We found that many words included in LIWC categories had different meanings depending on context (Section \ref{sec:word_usage}). For instance, a user who remarks ``this assignment is killing me'' is having a bad day, but one who remarks ``I'm thinking about killing myself'' needs immediate help. An analysis that counts occurrences of the word ``kill'' cannot distinguish these cases. Followup research could leverage contextualized word embeddings~\cite{Peters2018DeepCW,Devlin2019BERTPO} to identify which mentions of potentially harmful words demand attention. In addition, future work could complement our linguistically-driven approach by collecting and analyzing direct ratings from users~\cite{Wang2019WhenDW}.

Finally, future work could examine the long-term effects of online mental health conversations. When do these discussions lead to long-term mental health improvements, and what are the key linguistic traits that turn momentary improvement into long-term progress?



\section{Conclusion}

We conducted a large-scale analysis examining the effect of moderation on online mental health discussions. We found that moderation improved civility, supportiveness, and coherence. Our findings suggest that moderated mental health support conversations could be a scalable tool to combat the ongoing mental health crisis and set the stage for deeper explorations into the impact of moderator expertise, moderation style, and moderator training on conversation outcomes.


\section*{Acknowledgements} This research was supported in part by NSF grant IIS-1901386, Bill \& Melinda Gates Foundation (INV-004841), an Adobe Data Science Research Award, the Allen Institute Institute for Artificial Intelligence, and a Microsoft AI for Accessibility grant.



\appendix

\section{Potential Threats to Validity} \label{appx:validity_threats}
\subsection{Time Frame} \label{appx:time_frame}

\begin{figure}[b]
  \centering
  \includegraphics[width=0.8\columnwidth]{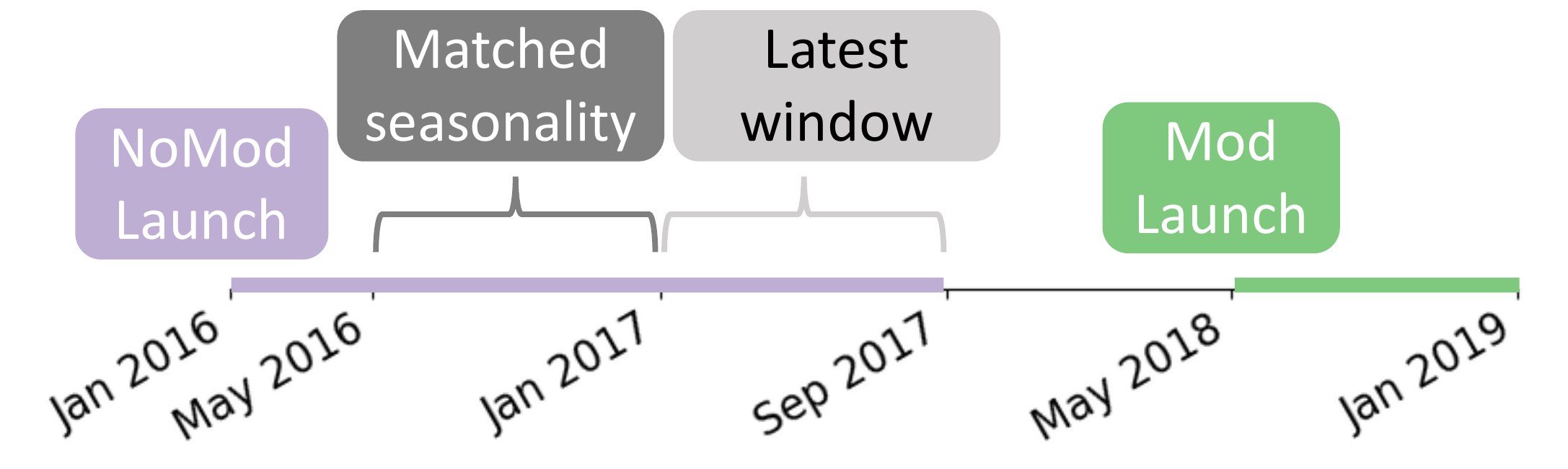}
  \caption{Time frames used for validity checks.}
  \label{fig:timeline}
\end{figure}

\begin{figure}[t]
  \centering
  \includegraphics[width=0.8\columnwidth]{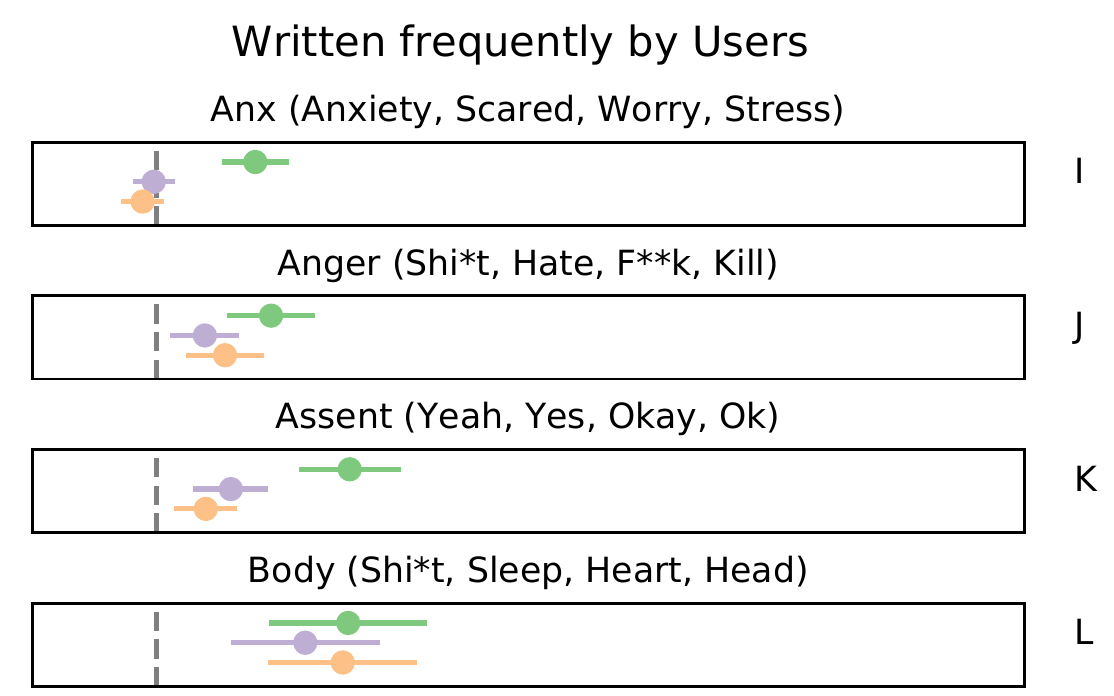}
  \caption{A replication of some rows from Figure \ref{fig:liwc_moderators_vs_users}, using only the \emph{matched-seasonality} subset of \nmod.}
  \label{fig:liwc_matched_season}
\end{figure}

To confirm that our results were robust to seasonality, we repeated our analyses using the \emph{matched-seasonality} subset of \nmod\ shown in Figure \ref{fig:timeline}, which runs for the same months as \ymod. Similarly, to verify that the results were not simply an artifact of changing behavioral norms or user demographics, we re-ran all analyses using the \emph{latest window} subset of \nmod\, which includes the final nine months of \nmod. The findings from the full dataset were robust to these perturbations. As one example, Figure \ref{fig:liwc_matched_season} shows an excerpt of Figure \ref{fig:liwc_moderators_vs_users}, re-created using only the \emph{matched-seasonality} subset of \nmod. The same trends are apparent here as in the full data.

\subsection{Discussion Topics} \label{appx:topics}

As a data-driven check that \ymod\ and \nmod\ centered on similar discussion topics, we fit an LDA topic model on all conversations, and computed the fraction of messages in \ymod\ and \nmod\ assigned to each topic. The results for a model fitted with 5 topics are shown in Table \ref{tbl:lda}. While the topic distributions are not statistically indistinguishable\footnote{A $\chi^2$ test rejects the null that \ymod\ and \nmod\ have identical topic distributions with $p < 0.001$.}, they are qualitatively quite similar; Topic 5 is the largest, followed by Topic 1, followed by 3 topics of roughly equal size. The same trends hold for a model with 10 topics.

\begin{table}[t]
  \centering

  \begin{tabular}{l r r r r r}
    \toprule
    & 1 & 2 & 3 & 4 & 5 \\
    \midrule
    Mod & 0.24 & 0.18 & 0.14 & 0.14 & 0.29 \\
    NoMod & 0.25 & 0.14 & 0.16 & 0.14 & 0.32 \\
    \bottomrule
  \end{tabular}
  \caption{Fraction of utterances assigned to each topic in a 5-topic LDA model.}
  \label{tbl:lda}
\end{table}


\subsection{Conversation Size} \label{appx:convo_size}

\begin{figure}[t]
  \centering
  \begin{subfigure}[t]{0.8\columnwidth}
    \includegraphics[width=\columnwidth]{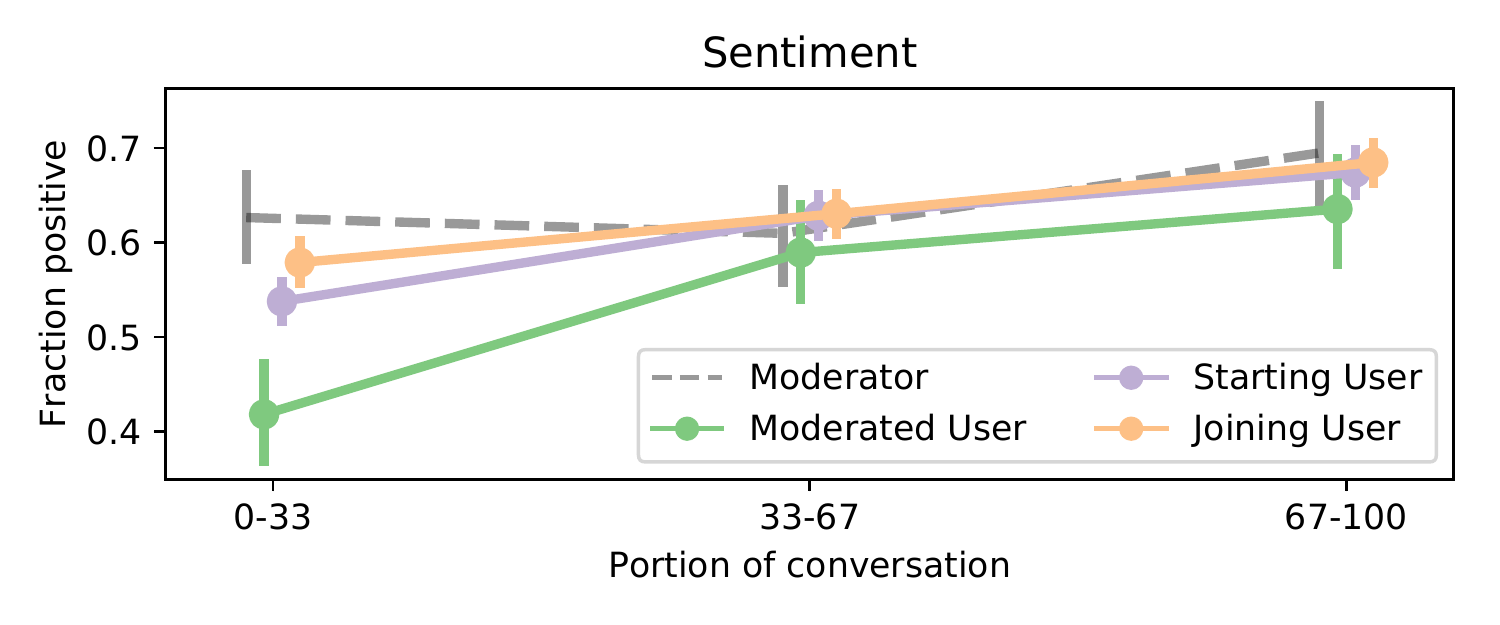}
    \subcaption{Sentiment change in one-on-one conversations.}
  \end{subfigure}
  \begin{subfigure}[t]{0.8\columnwidth}
    \includegraphics[width=\columnwidth]{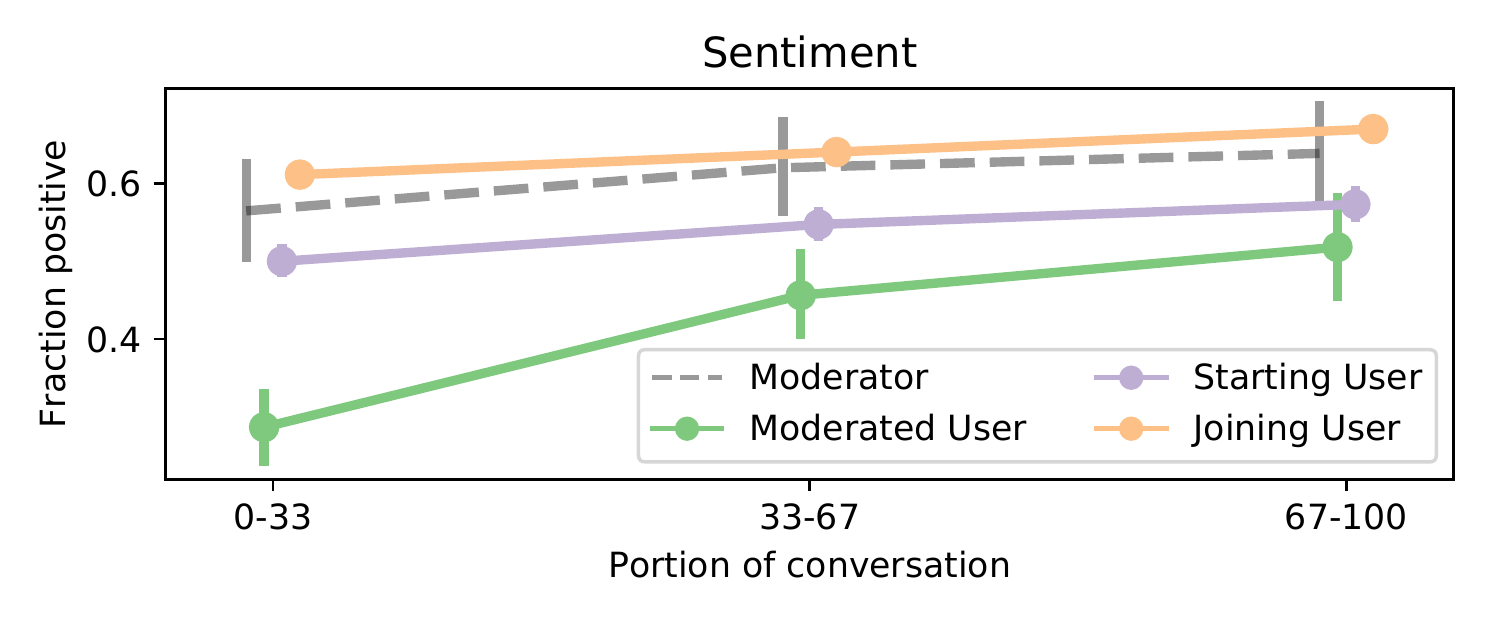}
    \subcaption{Sentiment change in multi-user conversations.}
  \end{subfigure}
  \caption{A replication of Figure \ref{fig:sentiment}, stratifying by number of conversation participants.}
  \label{fig:sentiment_stratified}
\end{figure}

We initially stratified our analyses by the number of conversation participants. We found that the same qualitative trends were present regardless of the number of participants, and we therefore collapsed the groups for our final analysis. As an example, Figure \ref{fig:sentiment_stratified} shows the sentiment changes from Figure \ref{fig:sentiment}, stratified into one-on-one conversations and conversations with at least 3 participants. While starting and joining users are more similar in one-on-one conversations, the same qualitative trends are apparent: user perspective improves over time, and these improvements are largest in moderated conversations.

\section{Moderator Activity} \label{appx:moderator_activity}

We stratified moderated conversations into three groups based on moderator activity, as measured by the fraction of messages in the conversation sent by the moderator. We refer to these three groups of messages as \ymod\ Active, \ymod\ Med, and \ymod\ Inactive. \ymod\ Active contains the top third of conversations with the highest fraction of moderator messages, \ymod\ Med has the middle third, and \ymod\ Inactive the lower third.

\begin{figure}[t]
  \centering
  \includegraphics[width=0.7\columnwidth]{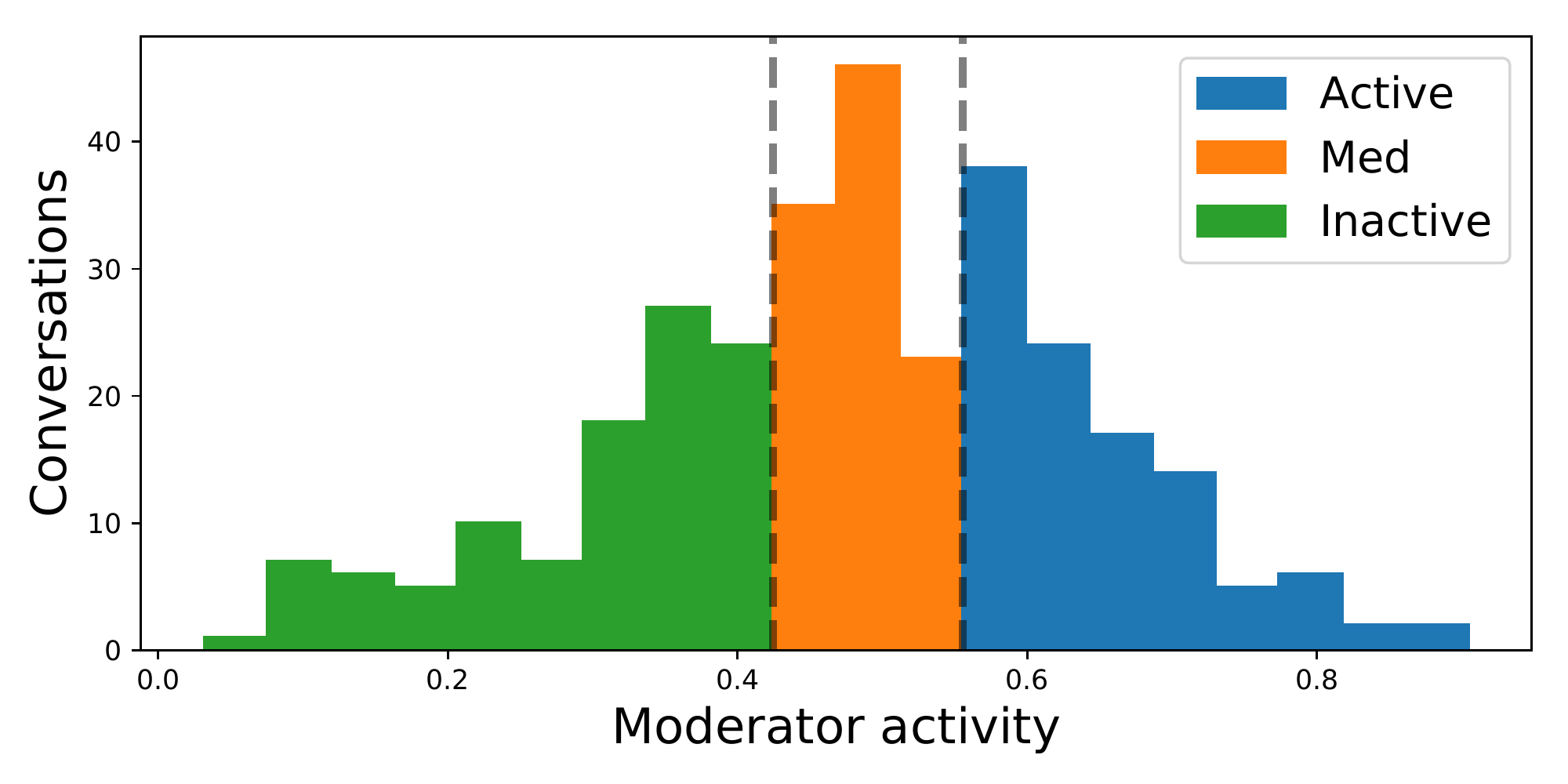}
  \caption{Distribution showing the fraction of messages in each conversation that were written by the moderator. Black dashed lines show 33rd and 67th quantiles.}
  \label{fig:moderator_involvement}
\end{figure}

The distribution of moderator activity is shown in Figure \ref{fig:moderator_involvement}. Activity varied widely, from very frequent posting to virtual absence. Moderator activity had an effect on the experiments presented in Section \ref{sec:relatedness}, but not elsewhere.

\section{LIWC Word Usage} \label{appx:math}

We describe the procedure used to compute the error bars on the relative differences in word usage shown in Figure 3. Let $g$ index the user groups and $k$ index the LIWC categories (e.g. ``Sad''). Denote the number of messages for each user group as $n_g$, and the collection of messages for group $g$ as $\{\mb_{g, i}\}_{i=1}^{n_g}$, where message $\mb_{g, i}$ is a sequence of tokens. For each LIWC category $k$, define $p_{g, i}^k$ to be the fraction of words in message $i$ of user group $g$ that match some word in LIWC category $k$. For instance, if the message were ``This makes me mad and upset'', and the LIWC category ``anger'' contained the words ``mad, upset, angry'', then $p_{g, i}^{\mathrm{anger}} = 2/6$. Then let
\begin{equation}
  \bar{p}_g^k = \frac{1}{n_g} \sum_{i=1}^{n_g} p_{g, i}^k,
\end{equation}
the average fraction of words matching category $k$ for messages from group $g$.
Similarly, let $\bar{p}_m^k$ be the average word fraction for messages sent by moderators. Define the relative change in word usage for LIWC category $k$ by user group $g$, relative to usage by moderators, as
\begin{equation}
  \Delta_g^k = \frac{(\bar{p}_g^k - \bar{p}_m^k)}{\bar{p}_m^k}
\end{equation}
Resample the user and moderator messages with replacement $n_{\mathrm{boot}} = 1000$ times, compute bootstrapped $\{\tilde{\Delta}_{j, g}^k\}_{j=1}^{n_\mathrm{boot}}$ on the resampled messages, and use the $2.5^{th}$ and $97.5^{th}$ quantiles as a 95\% confidence interval for $\Delta_g^k$.



\bibliography{supportiv}


\end{document}

%% file: gww-chars.tex
\newcommand{\mb}{{\mathbf{m}}}

